\documentclass[preprint,superscriptaddress,preprintnumbers,amsmath,amssymb,pra]{revtex4}
\usepackage{graphicx}

\begin{document}

\thispagestyle{empty}

\title{Low-temperature behavior of the Casimir-Polder free energy and
entropy for an atom interacting with graphene
}

\author{
G.~L.~Klimchitskaya}
\affiliation{Central Astronomical Observatory at Pulkovo of the
Russian Academy of Sciences, Saint Petersburg,
196140, Russia}
\affiliation{Institute of Physics, Nanotechnology and
Telecommunications, Peter the Great Saint Petersburg
Polytechnic University, Saint Petersburg, 195251, Russia}

\author{
V.~M.~Mostepanenko}
\affiliation{Central Astronomical Observatory at Pulkovo of the
Russian Academy of Sciences, Saint Petersburg,
196140, Russia}
\affiliation{Institute of Physics, Nanotechnology and
Telecommunications, Peter the Great Saint Petersburg
Polytechnic University, Saint Petersburg, 195251, Russia}
\affiliation{Kazan Federal University, Kazan, 420008, Russia}

\begin{abstract}
The analytic expressions for the free energy and entropy of the
Casimir-Polder interaction between a polarizable and magnetizable
atom and a graphene sheet are found in the limiting case of low
temperature. In so doing, the response of graphene to electromagnetic
fluctuations is described in the framework of the Dirac model by
means of the polarization tensor in (2+1)-dimensional space-time.
It is shown that the dominant contribution to the low-temperature
behavior is given by an explicit dependence of the polarization tensor
on temperature as a parameter. We demonstrate that the Lifshitz theory
of atom-graphene interaction satisfies the Nernst heat theorem, i.e.,
is thermodynamically consistent. On this basis possible reasons of
thermodynamic inconsistency arising for the Casimir-Polder and
Casimir interactions in the case of Drude metals are discussed.
The conclusion is made that although large thermal effect arising
in the Casimir interaction between Drude metals at short separations
should be considered as an artifact, the giant thermal effect
predicted for graphene systems is an important physical phenomenon
which awaits for its experimental observation.
\end{abstract}

\maketitle

\newcommand{\tp}{{\tilde{\Pi}}}
\newcommand{\rM}{{r_{\rm TM}}}
\newcommand{\rE}{{r_{\rm TE}}}
\newcommand{\orM}{{r_{\rm TM}^{(0)}}}
\newcommand{\orE}{{r_{\rm TE}^{(0)}}}
\newcommand{\zy}{{({\rm i}\zeta_l,y)}}
\newcommand{\ri}{{\rm i}}
\newcommand{\tg}{{\tilde{g}_l}}
\newcommand{\vF}{{\tilde{v}_F}}
\newcommand{\tB}{{\tilde{B}}}

\section{Introduction}

Physical phenomena known under a generic name of the Casimir-Polder
interaction refer to the fluctuation-induced forces acting between
polarizable and (or) magnetizable atoms (atomic systems) and
material surfaces.  These forces depend on the atomic and material
properties, on the atom-surface separation, and on the temperature.
At the shortest separations below a few nanometers they are of
nonrelativistic character and are often called the van der Waals
forces \cite{1}, whereas the relativistic generalization was
obtained by Casimir and Polder \cite{2} for the case of an
ideal-metal plane surface. Independently of separation, the
Casimir-Polder interaction is of entirely quantum nature. Similar
to the Casimir interaction, which refers to two macroscopic bodies
separated by a narrow gap, it is described by the Lifshitz theory
\cite{3}. In the framework of this theory, the Casimir-Polder
interaction was investigated for different atoms and surface
materials \cite{4,5,6,7,8,9,10,11}. Calculations of this kind are
useful for interpretation of experiments on Bose-Einstein
condensation \cite{12,13,14}, quantum reflection \cite{15,16,17}
and, e.g., for understanding of the resonance interaction of two
atoms near a boundary surface \cite{18}.

In the last few years much attention has been focused on the
Casimir-Polder interaction of different atoms with graphene and
graphene-coated substrates \cite{19,20,21,23,24,25,26,27,28,28a,29}.
Graphene is a two-dimensional sheet of carbon atoms packed in a
hexagonal lattice which possesses unusual electrical, optical and
mechanical properties \cite{30,31}. At low energies it is well
described by the Dirac model which assumes that graphene
quasiparticles are massless, obey a linear dispersion relation,
but move at the Fermi velocity $v_F \approx c$/300 in place of
the speed of light. As a result, the Casimir-Polder interaction
of atoms with a graphene sheet possesses the giant thermal effect
at short separations \cite{20} predicted earlier for the Casimir
force between two graphene sheets \cite{32}.

It has been known that large thermal effect at short separations
arises also in the Casimir interaction between two metallic
plates if the dielectric properties of metal at low frequencies
are described by the Drude model \cite{33,34}. Similar effect
arises in the Casimir-Polder force acting on an atom possessing
both electric polarizability and magnetic susceptibility when it
interacts with metallic plate described by the Drude model. For
the case of both nonmagnetic and magnetic metallic plates
described by the Drude model an existence of large thermal
effect in the Casimir force at short separations was
unambiguously excluded by many experiments
\cite{35,36,37,38,39,40,41,42,43}. On the theoretical side, it
was shown that the Lifshitz theory comes into conflict with the
Nernst heat theorem when the response of metals with perfect
crystal lattices to low-frequency electromagnetic fluctuations
is described by the Drude model. This was proven in different
geometries for the Casimir interaction between nonmagnetic
\cite{44,45,46,47,48} and magnetic \cite{49} metals and, very
recently, for the Casimir-Polder interaction of both polarizable
and magnetizable atoms interacting with metallic plate \cite{50}.
This raises a question of whether the theoretical description
of the Casimir-Polder interaction of atoms with graphene is
thermodynamically consistent. For the Casimir interaction
between two graphene sheets this fundamental question was
solved positively \cite{51}, but for an atom possessing both
the electric polarizability and magnetic susceptibility it
still remains unsolved.

In this paper, we investigate the low-temperature behavior of
the Casimir-Polder free energy and entropy for a polarizable
and magnetizable atom interacting with a graphene sheet. All
derivations are made in the framework of the Lifshitz theory,
and graphene is described by the Dirac model. The response of
graphene to electromagnetic fluctuations is found on the basis
of first principles of quantum electrodynamics at nonzero
temperature using the polarization tensor in (2+1)-dimensional
space-time. It is shown that both a summation over the discrete
Matsubara frequencies and an explicit temperature dependence of
the polarization tensor contribute to the Casimir-Polder free
energy and entropy. The behaviors of both the free energy and
entropy at low temperature are found analytically. In so doing,
the dominant contribution to them originates from an explicit
dependence of the polarization tensor on temperature. We
demonstrate that the Lifshitz theory of atom-graphene
interaction is in agreement with the Nernst heat theorem and,
thus, is thermodynamically consistent. On this basis some
conjectures concerning the reasons of inconsistency arising
when the Drude model is used are inferred.
Specifically, it is concluded that although large thermal effect
arising in the Casimir interaction between Drude metals should be
considered as an artifact, the giant thermal effect for graphene is
an important physical phenomenon which awaits for its experimental
observation.

The paper is organized as follows. In Sec.~II the general formalism for the
free energy of an atom-graphene interaction at low temperature is presented.
Section~III contains calculation of the contribution to the Casimir-Polder
free energy due to an implicit temperature dependence. The contribution due
to an explicit temperature dependence is found in Sec.~IV. In Sec.~V the
Nernst heat theorem for an atom interacting with graphene is proven and
some relevant problems are touched on. Section~VI contains our conclusions
and a discussion.

\section{The Casimir-Polder free energy for a polarizable and
magnetizable atom interacting with graphene at low temperature}

We consider an atom characterized by the dynamic electric polarizability
$\alpha(\omega)$ and magnetic susceptibility $\beta(\omega)$ separated
by a distance $a$ from a graphene sheet in thermal equilibrium at
temperature $T$. In this case the free energy  is
given by the Lifshits formula \cite{7,11}
\begin{eqnarray}
&&
{\cal F}(a,T)=-{k_BT}\sum_{l=0}^{\infty}{\vphantom{\sum}}^{\prime}
\int_{0}^{\infty}\!\!k_{\bot} dk_{\bot}q_le^{-2aq_l}
\nonumber \\
&&~~\times
\left\{\vphantom{\frac{\xi_l^2}{q_l^2c^2}}
2\left[\alpha_l\rM(\ri\xi_l,k_{\bot})+\beta_l\rE(\ri\xi_l,k_{\bot})
\right]\right.
\nonumber \\
&&~~~~~~\left.
-\frac{\xi_l^2}{q_l^2c^2}
(\alpha_l+\beta_l)
\left[\rM(\ri\xi_l,k_{\bot})+\rE(\ri\xi_l,k_{\bot})
\right]\right\}
.
\label{eq1}
\end{eqnarray}
\noindent
Here, $k_B$ is the Boltzmann constant, $k_{\bot}$ is the magnitude of the
projection of the wave vector on the plane of graphene,
$q_l^2=k_{\bot}^2+\xi_l^2/c^2$,
$\xi_l=2\pi k_BTl/\hbar$
with $l=0,\,1,\,2,\,\ldots$ are the Matsubara frequencies,
$\alpha_l=\alpha(\ri\xi_l)$, $\beta_l=\beta(\ri\xi_l)$, and
the prime on the summation sign  means that the term with $l=0$
has to be multiplied  by 1/2. The quantities $\rM$ and $\rE$ are the
reflection coefficients of electromagnetic fluctuations on graphene
with the transverse magnetic (TM) and transverse
electric (TE) polarizations.
Their explicit form is specified below.

Note that the free energy (\ref{eq1}) is an approximate expression obtained as
the first perturbation order in the small parameters $\alpha_l$ and $\beta_l$.
The nonperturbative generalization of the zero-temperature Casimir-Polder force
between an atom and an ideal-metal plane to the case $T\neq 0$
 was obtained only a few years ago \cite{52}. Very recently the
 nonperturbative generalization of Eq.~(\ref{eq1}) was also derived for the case
of any material plate \cite{53}. It was shown, however, that the exact and
perturbative free energies may differ for no more than 1\% and only at
$a<1$~nm \cite{53}. Taking into account that the Dirac model of graphene is
applicable at frequencies below approximately
$2~\mbox{eV}\approx 3.05\times 10^{15}$~rad/s, the formalism developed
in this section works good at atom-graphene separations $a>50$~nm.
In this separation region the perturbative free energy  (\ref{eq1}) is
indistinguishable from the exact one.

It is convenient to introduce the dimensionless variables
\begin{equation}
y=2q_la,\qquad
\zeta_l=\frac{2a\xi_l}{c}\equiv \tau l,
\label{eq2}
\end{equation}
\noindent
where
\begin{equation}
\tau=4\pi\frac{ak_BT}{\hbar c}=2\pi\frac{T}{T_{\rm eff}}
\label{eq3}
\end{equation}
\noindent
and the effective temperature for the Casimir effect is defined as
$k_BT_{\rm eff}=\hbar c/(2a)$.
In terms of these variables the free energy (\ref{eq1}) takes the form
\begin{eqnarray}
&&
{\cal F}(a,T)=-\frac{k_BT}{8a^3}\sum_{l=0}^{\infty}{\vphantom{\sum}}^{\prime}
\int_{\zeta_l}^{\infty}\!\!\!dy\,e^{-y}
\left\{
2y^2\left[\alpha_l\rM\zy+\beta_l\rE\zy
\right]\right.
\nonumber \\
&&~~~~~~\left.
-\zeta_l^2
(\alpha_l+\beta_l)
\left[\rM\zy+\rE\zy
\right]\right\}
.
\label{eq4}
\end{eqnarray}

The reflection coefficients $\rM$ and $\rE$ on  a graphene sheet have been expressed
via its polarization tensor in Ref.~\cite{54}.
Here we use an equivalent form for the reflection coefficients
\begin{eqnarray}
&&
r_{\rm TM}(\ri\zeta_l,y)=\frac{y\tilde{\Pi}_{00}(\ri\zeta_l,y)}{y
\tilde{\Pi}_{00}(\ri\zeta_l,y)+2(y^2-\zeta_l^2)},
\nonumber \\
&&
r_{\rm TE}(\ri\zeta_l,y)=-\frac{\tilde{\Pi}(\ri\zeta_l,y)}{\tilde{\Pi}(\ri\zeta_l,y)
+2y(y^2-\zeta_l^2)},
\label{eq5}
\end{eqnarray}
\noindent
where $\tp_{nm}$ with $n,\,m=0,\,1,\,2$ is the dimensionless polarization tensor
of graphene connected with the dimensional one, $\Pi_{nm}$,  by
$\tp_{nm}=2a\Pi_{nm}/\hbar$, the quantity $\tp$ is defined as
\begin{equation}
\tp\zy=(y^2-\zeta_l^2){\rm tr}\tp_{nm}-y^2\tp_{00}
\label{eq6}
\end{equation}
\noindent
and ${\rm tr}\tp_{nm}=\tp_{n}^{\,n}$ is the trace of the polarization tensor.

It is convenient to present the polarization tensor in the form
\begin{eqnarray}
&&
\tp_{00}\zy=\tp_{00}^{(0)}\zy+\Delta_T\tp_{00}\zy,
\nonumber \\
&&
\tp\zy=\tp^{(0)}\zy+\Delta_T\tp\zy,
\label{eq7}
\end{eqnarray}
\noindent
where $\tp_{00}^{(0)}$ and $\tp^{(0)}$ are found at $T=0$ but with continuous
dimensionless frequencies $\zeta$ replaced by the discrete Matsubara
frequencies $\zeta_l$ and $\Delta_T\tp_{00}$, $\Delta_T\tp$ have the meaning
of thermal corrections. The polarization tensor at $T=0$ has a very
simple form \cite{54,55}
\begin{eqnarray}
&&
\tp_{00}^{(0)}\zy=\frac{\pi\alpha(y^2-\zeta_l^2)}{\tg},
\label{eq8} \\
&&
\tp^{(0)}\zy={\pi\alpha(y^2-\zeta_l^2)}{\tg},
\nonumber
\end{eqnarray}
\noindent
where $\alpha=e^2/(\hbar c)$ is the fine structure constant,
\begin{equation}
\tg=\tg(y)=[\vF^2y^2+(1-\vF^2)\zeta_l^2]^{1/2}
\label{eq9}
\end{equation}
\noindent
and $\vF=v_F/c\approx 1/300$. Taking this into account, one can safely put
\begin{equation}
\tg\approx[\vF^2y^2+\zeta_l^2]^{1/2}.
\label{eq10}
\end{equation}

The thermal corrections to the polarization tensor can be presented in the
form valid only at the pure imaginary Matsubara frequencies \cite{54} and
over the entire plane of complex frequencies \cite{55,56}.
The latter form is used below. It is given by
\begin{eqnarray}
&&
\Delta_T\tilde{\Pi}_{00}(i\zeta_l,y)=\frac{8\alpha \tg}{\tilde{v}_F^2}
\int_{0}^{\infty}\!\!\frac{du}{e^{B_lu}+1}
\nonumber \\
&&~~~~~
\times\left\{1-\frac{1}{\sqrt{2}}\left[
\sqrt{(1+u^2)^2-4\frac{\tilde{v}_F^2(y^2-\zeta_l^2)u^2}{\tg^2}}+
1-u^2\right]^{1/2}\right\},
\label{eq11} \\
&&
\Delta_T\tilde{\Pi}(i\zeta_l,y)=\frac{8\alpha \tg}{\tilde{v}_F^2}
\int_{0}^{\infty}\frac{du}{e^{B_lu}+1}
\nonumber \\
&&~~~~~
\times\left\{
\vphantom{\frac{v_F^2k_{\bot}^2}{c^2\tilde{q}_l^2
\sqrt{(1+u^2)^2-4\frac{v_F^2k_{\bot}^2u^2}{c^2\tilde{q}_l^2}}}}
-{\zeta_l^2}+\frac{\tg^2}{\sqrt{2}}\left[
\sqrt{(1+u^2)^2-4\frac{\tilde{v}_F^2(y^2-\zeta_l^2)u^2}{\tg^2}}+
1-u^2\right]^{1/2}\right.
\nonumber \\
&&~~~~~
\times\left.\left[1-\frac{\tilde{v}_F^2(y^2-\zeta_l^2)}{\tg^2
\sqrt{(1+u^2)^2-4\frac{\tilde{v}_F^2(y^2-\zeta_l^2)u^2}{\tg^2}}}
\right]
\right\},
\nonumber
\end{eqnarray}
\noindent
where $B_l=\pi \tilde{g}_l/\tau$.

{}From Eq.~(\ref{eq11}) it is seen that $B_l\to\infty$ when $\tau\to 0$ and, thus,
\begin{equation}
\lim_{T\to 0}\Delta_T\tp_{00}\zy=\lim_{T\to 0}\Delta_T\tp\zy=0,
\label{eq12}
\end{equation}
\noindent
whereas, according to Eq.~(\ref{eq8}),
\begin{eqnarray}
&&
\lim_{T\to 0}\tp_{00}^{(0)}\zy=
\frac{\pi\alpha y}{\vF}\neq 0,
\nonumber \\
&&
\lim_{T\to 0}\tp^{(0)}\zy={\pi\alpha y^3}{\vF}\neq 0.
\label{eq13}
\end{eqnarray}
\noindent
Therefore, at sufficiently low $T$ one obtains
\begin{equation}
\frac{\Delta_T\tp_{00}\zy}{\tp_{00}^{(0)}\zy}\ll 1,
\qquad
\frac{\Delta_T\tp\zy}{\tp^{(0)}\zy}\ll 1.
\label{eq14}
\end{equation}

Substituting Eq.~(\ref{eq7}) in Eq.~(\ref{eq5}) and expanding up to the first
power in small parameters (\ref{eq14}), we find
\begin{eqnarray}
&&
\rM\zy=\orM\zy+\Delta_T\rM\zy,
\nonumber \\
&&
\rE\zy=\orE\zy+\Delta_T\rE\zy.
\label{eq15}
\end{eqnarray}
\noindent
Here, $r_{\rm TM(TE)}^{(0)}$ are the reflection coefficients at zero
temperature calculated at the pure imaginary Matsubara frequencies
\cite{55}
\begin{eqnarray}
&&
\orM\zy=\frac{\alpha\pi y}{\alpha\pi y+2\tg},
\nonumber \\
&&
\orE\zy=-\frac{\alpha\pi \tg}{\alpha\pi\tg+2y}.
\label{eq16}
\end{eqnarray}
\noindent
They are obtained by the substitution of Eq.~(\ref{eq8}) in place of
Eq.~(\ref{eq7}) in Eq.~(\ref{eq5}). The quantities $\Delta_Tr_{\rm TM(TE)}$ in
Eq.~(\ref{eq15}) have the meaning of the thermal corrections to the reflection
coefficients calculated up to the first order in parameters (\ref{eq14}).
They are given by
\begin{eqnarray}
&&
\Delta_T\rM\zy=\frac{2\alpha\pi y\tg}{(\alpha\pi y+2\tg)^2}
\frac{\Delta_T\tp_{00}\zy}{\tp_{00}^{(0)}\zy},
\nonumber \\
&&
\Delta_{T}\rE\zy=-\frac{2\alpha\pi y\tg}{(\alpha\pi\tg+2y)^2}
\frac{\Delta_T\tp\zy}{\tp^{(0)}\zy}.
\label{eq17}
\end{eqnarray}

It was shown \cite{51} that for sufficiently low temperatures,
satisfying the condition
\begin{equation}
k_BT\ll\frac{\hbar v_F}{2a}\equiv k_BT_{\rm eff}^{(g)},
\label{18}
\end{equation}
\noindent
where $T_{\rm eff}^{(g)}\neq T_{\rm eff}$ is one more effective temperature
for the Casimir effect in graphene systems, the dominant contributions to
the parameters (\ref{eq14}) at
$l\geq 1$ take the form
\begin{eqnarray}
&&
\frac{\Delta_T\tp_{00}\zy}{\tp_{00}^{(0)}\zy}=
\frac{48\zeta(3)}{\pi\tg^3}\left(\frac{T}{T_{\rm eff}}\right)^3,
\label{eq19} \\
&&
\frac{\Delta_T\tp\zy}{\tp^{(0)}\zy}=
\frac{96\zeta(3)}{\pi\tg^3}\left(\frac{T}{T_{\rm eff}}\right)^3
\left(\frac{3\zeta_l^2}{2\tg^2}-1\right).
\nonumber
\end{eqnarray}
Here, $\zeta(z)$ is the Riemann $\zeta$-function.

Finally, substitution of Eq.~(\ref{eq19}) in Eq.~(\ref{eq17}) results in the formulas
\begin{eqnarray}
&&
\Delta_{T}\rM\zy=
\frac{92\zeta(3)\alpha y}{\tg^2(\alpha\pi y+2\tg)^2}\left(\frac{T}{T_{\rm eff}}\right)^3,
\label{eq20} \\
&&
\Delta_{T}\rE\zy=
-\frac{192\zeta(3)\alpha y}{\tg^2(\alpha\pi\tg+2y)^2}
\left(\frac{T}{T_{\rm eff}}\right)^3
\left(\frac{3\zeta_l^2}{2\tg^2}-1\right).
\nonumber
\end{eqnarray}
\noindent
which are valid for any $l\geq 1$.
Equations (\ref{eq4}), (\ref{eq15}), (\ref{eq16}),  and (\ref{eq20})
are used below to find the low-temperature behavior of the Casimir-Polder
free energy and entropy (special attention will be given to the case
$l=0$).

\section{Contribution to the free energy due to implicit temperature
dependence}

Substituting  Eq.~(\ref{eq15}) in Eq.~(\ref{eq4}), one can present the
Casimir-Polder free energy as a sum of two contributions
\begin{equation}
{\cal F}(a,T)={\cal F}^{(1)}(a,T) +\Delta_T^{\!(2)}{\cal F}(a,T),
\label{eq21}
\end{equation}
\noindent
where both ${\cal F}^{(1)}$ and $\Delta_T^{\!(2)}{\cal F}$ have the
same form as Eq.~(\ref{eq4}), but the reflection coefficients $r_{\rm TM(TE)}$
are replaced with $r_{\rm TM(TE)}^{(0)}$ and $\Delta_{T}r_{\rm TM(TE)}$,
respectively.

In the lowest order with respect to the parameter $\zeta_l=\tau l$,
we can restrict our attention to the static electric polarizability and
magnetic susceptibility (note that the latter is essentially independent
of frequency for many atoms \cite{57}). If this is the case,
using Eq.~(\ref{eq4}),
the quantity ${\cal F}^{(1)}$ can be written in the form
\begin{equation}
{\cal F}^{(1)}(a,T)=-\frac{k_BT}{8a^3}
\sum_{l=0}^{\infty}{\vphantom{\sum}}^{\prime} \Phi(\tau l),
\label{eq22}
\end{equation}
\noindent
where
\begin{eqnarray}
&&
\Phi(\tau l)=
\int_{\tau l}^{\infty}\!\!\!dy\,e^{-y}
\left\{
\left[2y^2\alpha_0-(\tau l)^2(\alpha_0+\beta_0)\right]
\orM(\ri\tau l,y)\right.
\nonumber \\
&&~~~~~~\left. +
\left[2y^2\beta_0-(\tau l)^2(\alpha_0+\beta_0)\right]
\orE(\ri\tau l,y)\right\}
\label{eq23}
\end{eqnarray}
\noindent
and $\alpha_0=\alpha(0),{\ }\beta_0=\beta(0)$.

Note that the Casimir-Polder energy at zero temperature can be
represented in the same form \cite{34}
\begin{equation}
E(a)=-\frac{\hbar c}{32\pi a^4}
\int_{0}^{\infty}\!\!\!
d\zeta\Phi(\zeta),
\label{eq24}
\end{equation}
\noindent
where
\begin{eqnarray}
&&
\Phi(\zeta)=
\int_{\zeta}^{\infty}\!\!\!dy\,e^{-y}
\left\{
\left[2y^2\alpha_0-\zeta^2(\alpha_0+\beta_0)\right]
\orM(\ri\zeta,y)\right.
\nonumber \\
&&~~~~~~\left. +
\left[2y^2\beta_0-\zeta^2(\alpha_0+\beta_0)\right]
\orE(\ri\zeta,y)\right\}.
\label{eq25}
\end{eqnarray}
\noindent
It is seen that Eq.~(\ref{eq24}) is obtainable from
Eq.~(\ref{eq22}) in the case that
a summation over the discrete Matsubara frequencies is replaced with
an integration along the imaginary frequency axis in accordance to
the rule
\begin{equation}
k_BT\sum_{l=0}^{\infty}{\vphantom{\sum}}^{\prime} \to
\frac{\hbar c}{4\pi a}\int_{0}^{\infty}\!\!\!d\zeta.
\label{eq26}
\end{equation}

Thus, by applying the Abel-Plana formula \cite{34}
\begin{equation}
\sum_{l=0}^{\infty}{\vphantom{\sum}}^{\prime} \Phi(l)=
\int_{0}^{\infty}\!\!\Phi(t)dt+\ri
\int_{0}^{\infty}\!\frac{\Phi(\ri t)-\Phi(-\ri t)}{e^{2\pi t}-1}dt,
\label{eq27}
\end{equation}
\noindent
which is valid for any function analytic in the right half-plane,
one can represent the quantity ${\cal F}^{(1)}$ as
\begin{equation}
{\cal F}^{(1)}(a,T)=E(a)+\Delta_T^{\!(1)}{\cal F}(a,T),
\label{eq28}
\end{equation}
\noindent
where
\begin{equation}
\Delta_T^{\!(1)}{\cal F}(a,T)=
-\ri\frac{k_BT}{8a^3}
\int_{0}^{\infty}\!\frac{\Phi(\ri\tau t)-\Phi(-\ri\tau t)}{e^{2\pi t}-1}dt.
\label{eq29}
\end{equation}
\noindent
{}From Eqs.~(\ref{eq28}) and (\ref{eq29}) it is apparent that the thermal
correction $\Delta_T^{\!(1)}{\cal F}$ represents an implicit temperature
dependence of the free energy which arises from a summation over the
Matsubara frequencies in the contribution calculated with the zero-temperature
reflection coefficients.

Direct calculation using Eqs.~(\ref{eq25}), (\ref{eq16}), and (\ref{eq10})
results in
\begin{equation}
\Phi(\ri\tau t)-\Phi(-\ri\tau t)=-2\ri\tau^3t^3(C_{\rm TM}+C_{\rm TE}),
\label{eq30}
\end{equation}
\noindent
where
\begin{eqnarray}
&&
C_{\rm TM}=\frac{2\alpha\pi\alpha_0}{\vF(\alpha\pi+2\vF)^3}+
\frac{\alpha_0+3\beta_0}{3(\alpha\pi+2\vF)},
\nonumber \\
&&
C_{\rm TE}=\frac{\alpha\pi\vF\alpha_0}{2}-
\frac{\alpha\pi\beta_0}{2\vF}.
\label{eq31}
\end{eqnarray}
\noindent
Note that, when calculating $C_{\rm TE}$, we neglect by not only $\vF^2$,
but also by $\alpha\pi\vF$, as compared to unity. Note also that the next
contribution on the right-hand side of Eq.~(\ref{eq30}) is of the order of
$\tau^4\ln\tau$.

Substituting Eqs.~(\ref{eq30}) and (\ref{eq31}) in
Eq.~(\ref{eq29}), one obtains
\begin{eqnarray}
\Delta_T^{\!(1)}{\cal F}(a,T)&=&-\frac{k_BT}{4a^3}\tau^3
(C_{\rm TM}+C_{\rm TE})\int_{0}^{\infty}\!\!\frac{t^3\,dt}{e^{2\pi t}-1}
\nonumber \\
&=&-\frac{\pi^3(k_BT)^4}{15(\hbar c)^3}
(C_{\rm TM}+C_{\rm TE}).
\label{eq32}
\end{eqnarray}
\noindent
This result is of the same order in $T$ as for an atom interacting with
dielectric  plate \cite{34}.

The main contribution to Eq.~(\ref{eq32}) is given by the first term in the
coefficient $C_{\rm TM}$ defined in Eq.~(\ref{eq31}). Thus, if we assume that
$\alpha_0\sim\beta_0$, the first term in $C_{\rm TM}$ is more than the second one
and than $|C_{\rm TE}|$ by the factors of $5\times 10^5$ and $1.5\times 10^5$,
respectively. If we assume that $\alpha_0\gg\beta_0$, the same ratios are
equal to $\approx 2\times 10^6$ and $\approx 4\times 10^{11}$,
respectively.

\section{Contribution to the free energy due to explicit temperature
dependence}

We are coming now to the second contribution to the Casimir-Polder
free energy on the right-hand side of Eq.~(\ref{eq21}). It is obtained by a
replacement of the reflection coefficients $r_{\rm TM(TE)}$ in Eq.~(\ref{eq4})
with $\Delta_{T}r_{\rm TM(TE)}$ defined in Eq.~(\ref{eq17})
\begin{eqnarray}
&&
\Delta_T^{\!(2)}{\cal F}(a,T)=-\frac{k_BT}{8a^3}
\sum_{l=0}^{\infty}{\vphantom{\sum}}^{\prime}
\int_{\zeta_l}^{\infty}\!\!\!dy\,e^{-y}
\nonumber \\
&&~~~\times
\left\{2y^2
\left[\alpha_0\Delta_{T}\rM(\ri\zeta_l,y)+
\beta_0\Delta_{T}\rE(\ri\zeta_l,y)\right]
\right.
\nonumber \\
&&~~~~~~\left.
-\zeta_l^2(\alpha_0+\beta_0)
\left[\Delta_{T}\rM(\ri\zeta_l,y)+
\Delta_{T}\rE(\ri\zeta_l,y)\right]
\right\}.
\label{eq33}
\end{eqnarray}
\noindent
Here, as explained in Sec.~III, it is possible to restrict ourselves to
the case of static polarizability and susceptibility. This contribution
depends on the thermal correction to the polarization tensor of graphene
and takes into account its explicit dependence on the temperature as a
parameter. Here, we find the low-temperature behavior of Eq.~(\ref{eq33}).
For this purpose, we present Eq.~(\ref{eq33})
 as a sum of two terms
\begin{equation}
\Delta_T^{\!(2)}{\cal F}(a,T)=
\Delta_T^{\!(2)}{\cal F}_{(l\geq 1)}(a,T) +
\Delta_T^{\!(2)}{\cal F}_{(l=0)}(a,T)
\label{eq34}
\end{equation}
\noindent
and consider each of them separately.

We start from $\Delta_T^{\!(2)}{\cal F}_{(l\geq 1)}$ which includes all terms
of Eq.~(\ref{eq33}) except for the term with $l=0$.
In this case the thermal corrections to the reflection coefficients in the
lowest perturbation order are expressed by Eq.~(\ref{eq20}).
Substituting Eq.~(\ref{eq20}) in Eq.~(\ref{eq33}), one obtains
\begin{eqnarray}
&&
\Delta_T^{\!(2)}{\cal F}_{(l\geq 1)}(a,T)=
-\frac{12\zeta(3)\alpha k_BT}{a^3}
\left(\frac{T}{T_{\rm eff}}\right)^3
\sum_{l=1}^{\infty}
\int_{\tau l}^{\infty}\!\!\!dy\frac{ye^{-y}}{\tg^2}
\label{eq35} \\
&&~~\times
\left[\frac{2\alpha_0 y^2-(\tau l)^2(\alpha_0+
\beta_0)}{(\alpha\pi y+2\tg)^2}-
\frac{2\beta_0 y^2-(\tau l)^2(\alpha_0+
\beta_0)}{(\alpha\pi\tg+2y)^2}\left(
\frac{3\tau^2l^2}{\tg^2}-2\right)
\right].
\nonumber
\end{eqnarray}

Now we consider sufficiently low temperatures satisfying a condition
$\tau l\ll\vF y$ and expand the quantity $\tg^2$ defined in
Eq.~(\ref{eq10}) in
powers of the small parameter $(\tau l/\vF y)^2$ taking into account that
the major
contribution to the integrals in Eq.~(\ref{eq35}) is given by $y\sim 1$.
In so doing one can neglect by $3\tau^2l^2/(2\tg^2)$ as compared to unity.
We also take into account that $\alpha\pi\vF\approx 7.6\times 10^{-5}$ and,
thus, $\alpha\pi\tg\approx \alpha\pi\vF y\ll 2y$.
As a result, Eq.~(\ref{eq35}) can be rewritten as
\begin{eqnarray}
&&
\Delta_T^{\!(2)}{\cal F}_{(l\geq 1)}(a,T)=
-\frac{12\zeta(3)\alpha k_BT}{\vF^2a^3}
\left(\frac{T}{T_{\rm eff}}\right)^3
\sum_{l=1}^{\infty}
\int_{\tau l}^{\infty}\!\!\!dy\frac{e^{-y}}{y^3}
\label{eq36} \\
&&~~\times
\left[\frac{2\alpha_0 y^2-(\tau l)^2(\alpha_0+
\beta_0)}{(\alpha\pi +2\vF)^2}+
\frac{2\beta_0 y^2-(\tau l)^2(\alpha_0+
\beta_0)}{2}
\right].
\nonumber
\end{eqnarray}

After the change of integration variable $y=\tau lz$, we bring Eq.~(\ref{eq36})
to the form
\begin{eqnarray}
&&
\Delta_T^{\!(2)}{\cal F}_{(l\geq 1)}(a,T)=
-\frac{12\zeta(3)\alpha k_BT}{\vF^2a^3}
\left(\frac{T}{T_{\rm eff}}\right)^3
\nonumber \\
&&~~\times
\sum_{l=1}^{\infty}\left\{
\int_{1}^{\infty}\!\!\!dz\frac{e^{-\tau lz}}{z}
\left[\frac{2\alpha_0}{(\alpha\pi +2\vF)^2}+\beta_0\right]
\right.
\label{eq37} \\
&&~~~~\left.
-
\int_{1}^{\infty}\!\!\!dz\frac{e^{-\tau lz}}{z^3}(\alpha_0+
\beta_0)
\left[\frac{1}{(\alpha\pi +2\vF)^2}+\frac{1}{2}
\right]\right\}.
\nonumber
\end{eqnarray}

Now we make the summation first and calculate the integrals under a condition
$\tau\ll 1$ with the result
\begin{eqnarray}
&&
\Delta_T^{\!(2)}{\cal F}_{(l\geq 1)}(a,T)=
-\frac{12\zeta(3)\alpha k_BT}{\vF^2a^3}
\left(\frac{T}{T_{\rm eff}}\right)^3\frac{T_{\rm eff}}{T}
\label{eq38} \\
&&~~\times
\left\{
\frac{2\alpha_0}{(\alpha\pi +2\vF)^2}+\beta_0
-\frac{1}{2}(\alpha_0+\beta_0)
\left[\frac{1}{(\alpha\pi +2\vF)^2}+\frac{1}{2}
\right]\right\},
\nonumber
\end{eqnarray}
\noindent
or, equivalently,
\begin{equation}
\Delta_T^{\!(2)}{\cal F}_{(l\geq 1)}(a,T)=
-\frac{48\zeta(3)\alpha (k_BT)^3}{\vF^2(\hbar c)^2a}
(Q_1+Q_2),
\label{eq39}
\end{equation}
\noindent
where
\begin{equation}
Q_1=\frac{3\alpha_0-\beta_0}{2(\alpha\pi +2\vF)^2},
\quad
Q_2=-\frac{\alpha_0-3\beta_0}{4}.
\label{eq40}
\end{equation}
\noindent
It is easily seen that the major contribution to Eq.~(\ref{eq39}) is given
by the first term with a coefficient $Q_1$. For instance, if
$\alpha_0\sim\beta_0$ the coefficient $Q_1$ is larger than $Q_2$ by the
factor of 2300. If $\alpha_0\gg\beta_0$ one has $Q_1\approx 3400 Q_2$.

Comparing Eqs.~(\ref{eq32}) and (\ref{eq39}), we conclude that
 in the region of low temperatures an explicit
temperature dependence, originating from the Matsubara terms with $l\geq 1$,
is stronger than an implicit one.

Now we consider the term $\Delta_T^{\!(2)}{\cal F}_{(l=0)}$ on the right-hand
side of Eq.~(\ref{eq34}) which is equal to the zero-frequency contribution to
Eq.~(\ref{eq33}), i.e.,
\begin{eqnarray}
&&
\Delta_T^{\!(2)}{\cal F}_{(l=0)}(a,T)=
-\frac{k_BT}{8a^3}
\int_{0}^{\infty}\!\!\!dy\,e^{-y}y^2
\left[\alpha_0\Delta_{T}\rM(0,y)
\right.
\nonumber \\
&&~~~~~~~~
\left. +
\beta_0\Delta_{T}\rE(0,y)\right],
\label{eq41}
\end{eqnarray}
\noindent
where the thermal corrections to the reflection coefficients are
obtained from Eq.~(\ref{eq17})
\begin{eqnarray}
&&
\Delta_T\rM(0,y)=\frac{2\alpha\pi \vF}{(\alpha\pi +2\vF)^2}
\frac{\Delta_T\tp_{00}(0,y)}{\tp_{00}^{(0)}(0,y)},
\nonumber \\
&&
\Delta_{T}\rM(0,y)=-\frac{2\alpha\pi \vF}{(\alpha\pi\vF+2)^2}
\frac{\Delta_T\tp(0,y)}{\tp^{(0)}(0,y)}.
\label{eq42}
\end{eqnarray}
\noindent
Here, the polarization tensor at $T=0$, $\zeta_0=0$ is found
from Eq.~(\ref{eq8})
\begin{equation}
\tp_{00}^{(0)}(0,y)=\frac{\pi\alpha y}{\vF},
\quad
\tp^{(0)}(0,y)=\pi\alpha\vF y^3,
\label{eq43}
\end{equation}
\noindent
and the thermal correction to it at $\zeta_0=0$ from Eq.~(\ref{eq11})
\begin{eqnarray}
&&
\Delta_T\tp_{00}(0,y)=\frac{8\alpha y}{\vF}
\int_{0}^{\infty}\!\!\frac{du}{e^{B_0u}+1}(1-\sqrt{1-u^2}),
\nonumber \\
&&
\Delta_T\tp(0,y)=-{8\alpha y^3}{\vF}
\int_{0}^{\infty}\!\!\frac{du}{e^{B_0u}+1}\frac{u^2}{\sqrt{1-u^2}},
\label{eq44}
\end{eqnarray}
\noindent
where $B_0=\pi\vF y/\tau$.

It is more convenient to rearrange Eq.~(\ref{eq44}) to an equivalent form
\cite{56}
\begin{eqnarray}
&&
\Delta_T\tp_{00}(0,y)=\frac{32\alpha ak_BT}{\hbar c\vF^2}
\int_{0}^{1}\!\!\!dx\ln\left[1+e^{-\tB y\sqrt{x(1-x)}}\right],
\nonumber \\
&&
\Delta_T\tp(0,y)=-{16\alpha{\vF} y^3}
\int_{0}^{1}\!\!\frac{\sqrt{x(1-x)}dx}{e^{\tB y\sqrt{x(1-x)}}+1},
\label{eq45}
\end{eqnarray}
\noindent
where $\tB=2B_0/y=T_{\rm eff}^{(g)}/T$.

Substituting Eqs.~(\ref{eq43}) and (\ref{eq45}) in
Eqs.~(\ref{eq42}) and (\ref{eq41}),
one obtains
\begin{eqnarray}
&&
\Delta_T^{\!(2)}{\cal F}_{(l=0)}(a,T)=
-\frac{4\alpha k_BT}{a^3}
\left\{
\alpha_0\frac{2ak_BT}{\hbar c(\alpha\pi+2\vF)^2}
\right.
\nonumber \\
&&~~~
\times\int_{0}^{\infty}\!\!\!dy\,e^{-y}y
\int_{0}^{1}\!\!\!dx\ln\left[1+e^{-\tB y\sqrt{x(1-x)}}\right]
\label{eq46} \\
&&~~~~
\left. +
\beta_0\frac{\vF}{4}\int_{0}^{\infty}\!\!\!dy\,e^{-y}y^2
\int_{0}^{1}\!\!\frac{\sqrt{x(1-x)}dx}{e^{\tB y\sqrt{x(1-x)}}+1}
\right\},
\nonumber
\end{eqnarray}
\noindent
where we have neglected by $\pi\alpha\vF$, as compared to 2, in the
second line of Eq.~(\ref{eq42}). Taking into account that
\begin{equation}
\frac{\sqrt{x(1-x)}}{e^{\tB y\sqrt{x(1-x)}}+1}=
-\frac{1}{\tB}\frac{d}{dy}\ln\left[1+e^{-\tB y\sqrt{x(1-x)}}\right],
\label{eq47}
\end{equation}
\noindent
and integrating by parts with respect to $y$, the second term of
Eq.~(\ref{eq46})
can be rewritten in the form
\begin{equation}
\beta_0\frac{\vF}{4\tB}\int_{0}^{1}\!\!\!dx
\int_{0}^{\infty}\!\!\!dy(2-y)ye^{-y}
\ln\left[1+e^{-\tB y\sqrt{x(1-x)}}\right].
\label{eq48}
\end{equation}

Now  we represent the logarithms in both the first and second terms of
Eq.~(\ref{eq46}) as the power series in $\exp[-\tB y\sqrt{x(1-x)}]$ and
integrate with respect to $y$
\begin{eqnarray}
&&
\Delta_T^{\!(2)}{\cal F}_{(l=0)}(a,T)=
-\frac{8\alpha(k_BT)^2}{a^2\hbar c}
\left\{
\vphantom{\left[\frac{1}{[\tB \sqrt{x(1-x)}]^3}\right]}
\frac{\alpha_0}{(\alpha\pi+2\vF)^2}
\right.
\nonumber \\
&&~~~\times
\sum_{n=1}^{\infty}
\frac{(-1)^{n-1}}{n}
\int_{0}^{1}\!\!\frac{dx}{[1+n\tB \sqrt{x(1-x)}]^2}
\label{eq49} \\
&&~~~~
+\frac{\beta_0}{2}\sum_{n=1}^{\infty}
\frac{(-1)^{n-1}}{n}
\int_{0}^{1}\!\!\!dx\left[\frac{1}{[1+n\tB \sqrt{x(1-x)}]^2}
\right.
\nonumber \\
&&~~~~~~~~
\left.\left.
-\frac{1}{[1+n\tB \sqrt{x(1-x)}]^3}\right]
\right\}.
\nonumber
\end{eqnarray}

It is convenient to introduce one more small parameter
$b_n=1/(n\tB)=T/(nT_{\rm eff}^{(g)})$ and define the integrals
\begin{equation}
I_k(b_n)=
\int_{0}^{1}\!\!\frac{dx}{[b_n + \sqrt{x(1-x)}]^k},
\label{eq50}
\end{equation}
\noindent
where $k=2,\,3$ and $b_n<1/2$. Then Eq.~(\ref{eq49}) takes the form
\begin{eqnarray}
&&
\Delta_T^{\!(2)}{\cal F}_{(l=0)}(a,T)=
-\frac{8\alpha(k_BT)^2}{a^2\hbar c}
\left(\frac{T}{T_{\rm eff}^{(g)}}\right)^2
\nonumber \\
&&~~~\times\left\{
\frac{\alpha_0}{(\alpha\pi+2\vF)^2}
\sum_{n=1}^{\infty}
\frac{(-1)^{n-1}}{n}
I_2(b_n)\right.
\label{eq51} \\
&&~~~~
\left. +
\frac{\beta_0}{2}\sum_{n=1}^{\infty}
\frac{(-1)^{n-1}}{n^3}
\left[I_2(b_n)-b_nI_3(b_n)\right]
\right\}.
\nonumber
\end{eqnarray}

Direct calculation results in
\begin{eqnarray}
&&
I_2(b_n)=-\frac{4}{1-4b_n^2}\left[
\vphantom{\left(\ln\frac{\sqrt{1-4b_n^2}}{\sqrt{1-4b_n^2}}\right)}
1+\frac{1}{\sqrt{1-4b_n^2}}\right.
\label{eq52} \\
&&~~
\times\left.
\left(\ln\frac{1-\sqrt{1-4b_n^2}}{1+\sqrt{1-4b_n^2}}
-\ln\frac{1+2b_n-\sqrt{1-4b_n^2}}{1+2b_n+\sqrt{1-4b_n^2}}\right)
\right],
\nonumber \\
&&
I_3(b_n)=\frac{4}{(1-4b_n^2)^2}\left[
\vphantom{\left(\ln\frac{\sqrt{1-4b_n^2}}{\sqrt{1-4b_n^2}}\right)}
\frac{1+8b_n^2}{2b_n}+
\frac{6b_n}{\sqrt{1-4b_n^2}}\right.
\nonumber \\
&&~~
\times\left.
\left(\ln\frac{1-\sqrt{1-4b_n^2}}{1+\sqrt{1-4b_n^2}}
-\ln\frac{1+2b_n-\sqrt{1-4b_n^2}}{1+2b_n+\sqrt{1-4b_n^2}}\right)
\right].
\nonumber
\end{eqnarray}
\noindent
Expanding Eq.~(\ref{eq52}) in powers of $b_n$ one obtains
\begin{eqnarray}
&&
I_2(b_n)=-4\ln{b_n}+O(b_n^0),
\label{eq53} \\
&&
I_3(b_n)=\frac{2}{b_n}+24b_n\ln{b_n}+O(b_n^0).
\nonumber
\end{eqnarray}
\noindent
Substituting these results in Eq.~(\ref{eq51}) and finding main contributions
to the sums in $n$, we arrive at
\begin{equation}
\Delta_T^{\!(2)}{\cal F}_{(l=0)}(a,T)=
\frac{96\alpha\zeta(3)(k_BT)^4}{\vF^2(\hbar c)^3}
(R_1+R_2)\ln\frac{2ak_BT}{\hbar\vF c},
\label{eq54}
\end{equation}
\noindent
where
\begin{equation}
R_1=\frac{\alpha_0}{(\alpha\pi+2\vF)^2},
\quad
R_2=\frac{\beta_0}{4}.
\label{eq55}
\end{equation}
\noindent
The major contribution to Eq.~(\ref{eq54}) is given by the first term with the
coefficient $R_1$. For $\alpha_0\sim \beta_0$ $R_1$ is larger than $R_2$
by the factor of $\approx 1100$ and all the more if $\alpha_0\gg \beta_0$.

As is seen in Eq.~(\ref{eq54}), with decreasing $T$ down to zero temperature
$\Delta_T^{\!(2)}{\cal F}_{(l=0)}$ becomes greater than
$\Delta_T^{\!(1)}{\cal F}$, determined by the implicit dependence on the
temperature, but less than
$\Delta_T^{\!(2)}{\cal F}_{(l\geq 1)}$ originating from the explicit
temperature dependence of all Matsubara terms with nonzero frequency.

\section{Low-temperature behavior of the free energy and entropy}

In Secs.~III and IV we have found the low-temperature behavior of all
contributions to the Casimir-Polder free energy. According to
Eqs.~(\ref{eq21}), (\ref{eq28}), and (\ref{eq34}), the free energy is given by
\begin{equation}
{\cal F}(a,T)=E(a)+\Delta_T^{\!(1)}{\cal F}(a,T)+
\Delta_T^{\!(2)}{\cal F}_{(l\geq 1)}(a,T)+
\Delta_T^{\!(2)}{\cal F}_{(l=0)}(a,T),
\label{eq56}
\end{equation}
\noindent
where $\Delta_T^{\!(1)}{\cal F}$ in Eq.~(\ref{eq32}) presents the implicit
low-temperature behavior originating exclusively from a summation over
the Matsubara frequencies whereas $\Delta_T^{\!(2)}{\cal F}_{(l\geq 1)}$
and $\Delta_T^{\!(2)}{\cal F}_{(l=0)}$ found in
Eqs.~(\ref{eq39}) and (\ref{eq54}),
respectively, are determined by the explicit dependence of the polarization
tensor on temperature as a parameter. As is seen from Eqs.~(\ref{eq32}),
(\ref{eq39}), and (\ref{eq54}), with decreasing temperature the major
contribution is given by  $\Delta_T^{\!(2)}{\cal F}_{(l\geq 1)}$.
Thus, from  Eqs.~(\ref{eq39}) and (\ref{eq56}) one can conclude that
\begin{equation}
{\cal F}(a,T)=E(a)
-\frac{48\zeta(3)\alpha (k_BT)^3}{\vF^2(\hbar c)^2a}
(Q_1+Q_2),
\label{eq57}
\end{equation}
\noindent
where the coefficients $Q_1$ and $Q_2$ are given in Eq.~(\ref{eq40}).

{}From Eq.~(\ref{eq57}) one obtains the low-temperature behavior of the
Casimir-Polder entropy
\begin{equation}
S(a,T)=-\frac{\partial{\cal F}(a,T)}{\partial T}=
\frac{144\zeta(3)\alpha k_B(k_BT)^2}{\vF^2(\hbar c)^2a}
(Q_1+Q_2).
\label{eq58}
\end{equation}
\noindent
{}From Eq.~(\ref{eq54}) we find that the next term in the low-temperature
behavior of the entropy is of the order of
\begin{equation}
k_BR_1\frac{(k_BT)^3}{(\hbar c)^3}\ln\frac{2ak_BT}{\hbar\vF c}.
\label{eq59}
\end{equation}

Equation (\ref{eq58}) allows to make a conclusion that the Casimir-Polder
entropy is positive and goes to zero with vanishing temperature in
accordance with the third law of thermodynamics, the Nernst heat theorem.
This means that the Lifshitz theory of atom-graphene interaction is
thermodynamically consistent if the response of graphene to a fluctuating
field is described by the polarization tensor in the framework of the
Dirac model.

This fundamental result returns us to the problem discussed in Sec.~I,
i.e., why the Lifshitz theory of the Casimir and Casimir-Polder interaction
violates the Nernst heat theorem and is inconsistent with the measurement
data of several experiments if the low-frequency electromagnetic response
of metals is described by the well tested
under ordinary conditions Drude model taking into account
the relaxation properties of free charge carriers.

In connection with this, it is significant that the response of graphene
to electromagnetic fluctuations is described by the polarization tensor
on the basis of first principles of quantum electrodynamics at
nonzero temperature. This description is in full agreement with all
fundamental demands, such as causality, and satisfies the Kramers-Kronig
relations \cite{62}. By contrast, the Drude dielectric permittivity,
\begin{equation}
\varepsilon_D(\omega)=1-\frac{\omega_p^2}{\omega[\omega+\ri\gamma(T)]},
\label{eq60}
\end{equation}
\noindent
where $\omega_p$ is the plasma frequency and $\gamma(T)$ is the relaxation
parameter, is of entirely phenomenological character.
Although it provides an adequate description of the electrical conductivity
and optical properties of metals and satisfies the Kramers-Kronig
relations at nonzero temperature, the problem arises in the limiting case
of vanishing temperature.

The point is that for metals with perfect crystal lattices $\gamma(T)$
vanishes when $T$ goes to zero \cite{63}. In this case \cite{64}
\begin{equation}
\lim_{\gamma\to 0}\varepsilon_D(\omega)=
1-\frac{\omega_p^2}{\omega^2}+
\ri\frac{\omega_p^2}{\omega}\pi\delta(\omega),
\label{eq61}
\end{equation}
\noindent
where $\delta(\omega)$ is the Dirac $\delta$ function. This means that in the
limit of zero temperature the Drude dielectric permittivity cannot be continued
to the upper half plane of complex frequency and its imaginary part cannot be
obtained from its real part by means of the Kramers-Kronig relation \cite{65}.
Thus, in the limit of zero temperature the Drude model violates the principle
of causality and cannot be used as a dielectric permittivity. This gives an
insight into why the Lifshitz theory combined with the Drude model is in
conflict with the Nernst heat theorem. Note that if the plasma model is used
for description of the electromagnetic response
of a metal, i.e., $\gamma(T)$ in Eq.~(\ref{eq60}) is put equal
to zero from the outset, the Nernst heat theorem for the Casimir and Casimir-Polder
entropy is satisfied, as well as the Kramers-Kronig relations in the form valid
for functions possessing the second-order pole at zero frequency \cite{66}.

\section{Conclusions and discussion}

In the foregoing, we have analyzed the thermodynamic consistency of the Lifshitz
theory used for description of the Casimir-Polder interaction between a
polarizable and magnetizable atom and a graphene sheet. In so doing,
the response of graphene to electromagnetic fluctuations was described by the
polarization tensor in (2+1)-dimensional space-time in the framework of
quantum electrodynamics at nonzero temperature. We have found  analytic
expressions for the Casimir-Polder free energy and entropy at low temperature.
For this purpose the thermal correction to the Casimir-Polder energy was
represented as a sum of three contributions. The first of them originates
from a summation on the pure imaginary Matsubara frequencies and two other
from an explicit dependence of the polarization tensor on temperature as a
parameter. It was shown that the dominant contribution to the free energy
and entropy at low temperature is given by an explicit temperature
dependence contained in the nonzero-frequency terms of the Lifshitz formula.

Using the obtained analytic results, it was demonstrated that the
Casimir-Polder entropy of a polarizable and magnetizable atom interacting with
a graphene sheet satisfies the Nernst heat theorem. Thus, the Lifshitz theory
of an atom interacting with graphene
is thermodynamically consistent.
This fact was correlated with a violation of the Nernst theorem in the Lifshitz
theory of Casimir and Casimir-Polder interactions in  case that the
plate metal is described by the phenomenological Drude model. Special attention
was paid to the fact that the Drude dielectric function with vanishing relaxation
parameter ceases to be an analytic function in the upper half-plane of complex
frequency and violates the principle of causality. This can be considered as
a possible reason of thermodynamic inconsistency. By contrast, the response
of graphene to electromagnetic field is described on the basis of first
principles of quantum field theory and is in agreement
with the Kramers-Kronig relations
for all values of parameters.

On the basis of this discussion we conclude that large thermal effect at short
separations predicted by the Lifshitz theory for Drude metals and already excluded
experimentally should be considered as an artifact. As to the giant thermal effect
in the Casimir and Casimir-Polder interactions for graphene, this is an important
physical phenomenon which awaits for experimental observation. Two realistic
possibilities on how to observe this effect
have been proposed recently \cite{67,68,69} making its discovery in near future
very likely.

\section*{Acknowledgments}
The work of V.M.M. was partially supported by the Russian
Government
Program of Competitive Growth of Kazan Federal University.

\end{document}